\documentclass[pra,twocolumn,showpacs,amsmath,amssymb]{revtex4}
\usepackage{graphicx}
\usepackage{bm}
\usepackage{amsmath,amssymb}
\begin{document}

\title{Strong entanglement causes low gate fidelity
in inaccurate one-way quantum computation}

\author{Tomoyuki Morimae}
\email{morimae@gmail.com}
\affiliation{
Laboratoire Paul Painlev\'e, Universit\'e Lille 1,
F-59655 Villeneuve d'Ascq Cedex, France
}
\date{\today}
            
\begin{abstract}
We study how entanglement among the register
qubits affects the gate fidelity in the one-way
quantum computation if a measurement is inaccurate.
We derive an inequality which shows that
the mean gate fidelity is upper bounded by
a decreasing function of the magnitude of the error of the 
measurement and
the amount of the entanglement
between the measured qubit and other register qubits.
The consequence of this inequality is that,
for a given amount of entanglement, which is theoretically calculated
once the algorithm is fixed, 
we can estimate from this inequality
how small the magnitude of the error should be
in order not to make the gate fidelity below a threshold,
which is specified by a technical requirement in a particular experimental setup
or by the threshold theorem of the fault-tolerant quantum computation.
\end{abstract}
\maketitle  

\section{Introduction}
\label{introduction}
The one-way quantum computation~\cite{cluster}
is a novel scheme of quantum computation often contrasted with
the traditional circuit model of quantum computation~\cite{Nielsen}. 
It is believed to be
one of the most promising approaches to 
the realization of scalable quantum computers, 
and indeed, 
small-size cluster states 
have already been created
in laboratories~\cite{Furusawa}. 
Some important quantum algorithms,
such as Deutsch's algorithm~\cite{Tame}
and
Grover's search algorithm~\cite{Walther},
have also been demonstrated experimentally.

One great advantage of the one-way quantum computation over
the circuit model is that  
the preparation 
of the resource (=entanglement) and
the consumption of it are clearly separated with each other.
This fact has prompted many researchers to explore
lower-bounds or upper-bounds for
the proper amount of resource entanglement for
the one-way quantum computation~\cite{Nest,Gross,Bremner,novelcluster}.
For example, it was shown~\cite{Nest} that a certain amount of
entanglement is necessary for any universal resource state for the
one-way quantum computation.
On the other hand, it was shown~\cite{Gross,Bremner} that
a state having too much entanglement is useless for 
the one-way quantum computation. 
These important results and
further research based on them will ultimately enable us
to pin down the exact amount of resource entanglement 
which is neither
too small nor too large
for the one-way quantum computation. 

If the proper amount of resource entanglement for
the one-way quantum computation is determined,
the next goal is to clarify how such proper entanglement affects the gate
fidelity of the
one-way quantum computation. Because
a highly-entangled state is often fragile~\cite{p,macroent,tim,magnon},
we cannot make the most of the power of entanglement
if the one-way quantum computation itself is unstable.
Of course, a one-way quantum computer is, like the circuit model of
a quantum computer, finally
stabilized to some extent by embedding a quantum error-correcting
code as shown in Ref.~\cite{Dawson}. 
However, it is still very important to investigate
the stability of a bare one-way quantum computer 
for several reasons~\cite{Tame3}. First, it gives valuable feedback for the 
study of general fault-tolerant schemes.
Second, it helps 
the development of ``made-to-measure" error-correcting codes.
Third, what experimentalists are now interested in is not the
gigantic fully-fledged quantum computer but a 
bare elementary gate between a couple of qubits.
Finally, and most importantly,
although the stability of the final result of the computation is guaranteed by
the threshold theorem,
we must verify the stability of each gate independently,
because the crucial assumption of the threshold theorem is that
the fidelity of each gate is larger than a certain
threshold~\cite{Dawson}. 

In this Rapid Communication, we study how the gate fidelity of the
one-way quantum computation is affected by the amount of
entanglement between the measured
qubit and other register qubits
if the measurement is inaccurate
in the sense that the direction to which the qubit is projected
is slightly deviated from the ideal one.
As the resource state having a proper amount of entanglement,
we adopt the cluster state~\cite{other}. 
Our main result is 
\begin{eqnarray*}
F\le 1-S\sin^2\frac{\epsilon}{2},
\end{eqnarray*}
which shows that the
mean gate fidelity $F$ $(0\le F\le 1)$ is upper bounded by the decreasing function of
the amount $S$ $(0\le S\le 1)$ of entanglement 
and the magnitude $\epsilon$ $(0\le \epsilon\ll 1)$ of the deviation.
The main consequence of this inequality is that,
for a given amount $S$ of entanglement, which is theoretically calculated
once the algorithm is fixed and is 
{\it often very large} (see Refs.~\cite{Jozsa,Ukena,visual,macroent} 
and Sec.~\ref{discussion}),
we can estimate from this inequality
how small $\epsilon$ should be in order not to
make the gate fidelity $F$ below a threshold,
which is specified by an experimentalist implementing the one-way
quantum computation on his/her particular experimental instruments
or by the threshold theorem of the fault-tolerant quantum computation.

\section{Setups}
\begin{figure}[htbp]
\begin{center}
\includegraphics[width=0.4\textwidth]{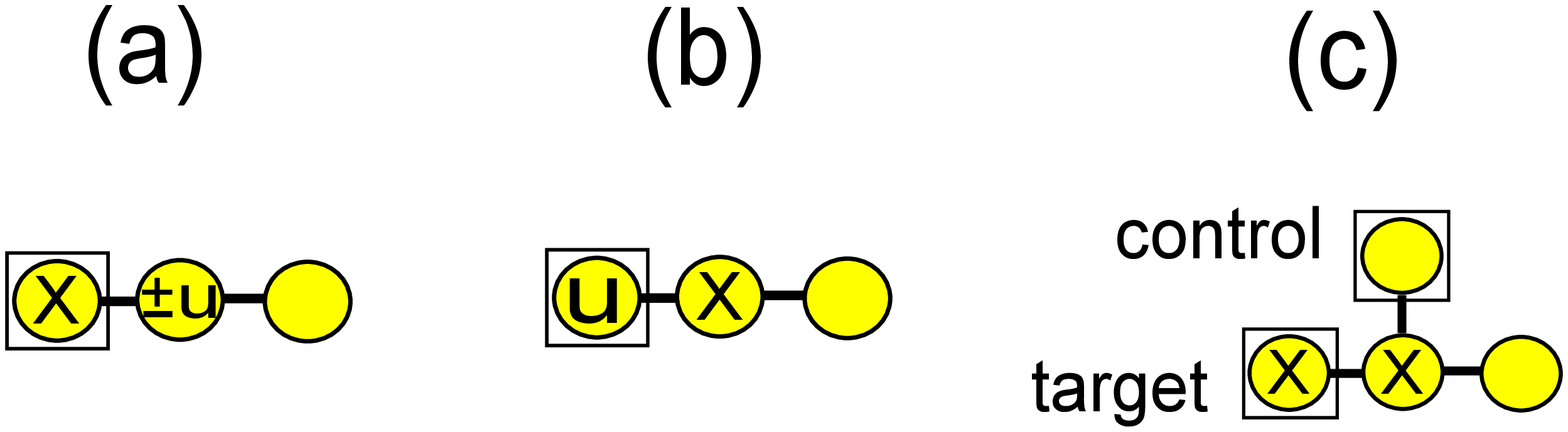}
\end{center}
\caption{(Color online.) 
Circles represent qubits,
bonds represent the controlled-Z (C-Z) interaction 
$|0\rangle\langle0|\otimes\hat{1}+|1\rangle\langle1|\otimes\hat{Z}$, and
squares represent the input state.
$X$ represents the measurement in $\hat{X}$-basis. 
$\pm u$ represents the adaptive measurement
in $(\cos u\hat{X}\mp\sin u\hat{Y})$-basis 
according to the result of the previous measurement $\pm$, respectively.
$u$ represents the measurement 
in $(\cos u\hat{X}-\sin u\hat{Y})$-basis.
Output states are modified according to the 
measurement history~\cite{debbie}.
(a): The single-qubit rotation 
$e^{-i\frac{u}{2}\hat{X}}$ by $u$
about $x$-axis.
(b): The single-qubit rotation 
$e^{-i\frac{u}{2}\hat{Z}}$ by $u$ about $z$-axis.
(c): The C-NOT gate.
} 
\label{gates}
\end{figure}

Before showing our main result, some setups are necessary.
As a universal set of quantum gates, we adopt the set of
single-qubit rotations about $x$-axis and $z$-axis,
and the controlled-NOT (C-NOT) gate between
two qubits~\cite{Nielsen}.
This is a universal gate set, since,
according to the Euler decomposition,
any single-qubit rotation can be written
as a combination of these two types of rotations.
We denote 
Pauli's $x$,$y$, and $z$ operators acting
on $i$ th qubit by $\hat{X}_i$, $\hat{Y}_i$, and $\hat{Z}_i$,
respectively. We also define eigenvectors of $\hat{X}_i$ and $\hat{Z}_i$ by
$\hat{X}_i|\pm\rangle_i=\pm|\pm\rangle_i$
and $\hat{Z}_i|z\rangle_i=(-1)^z|z\rangle_i$ $(z=0,1)$,
respectively.
Let us remind~\cite{debbie} that
the single-qubit rotation $e^{-i\frac{u}{2}\hat{X}}$ by $u$ 
about $x$-axis, 
the single-qubit rotation $e^{-i\frac{u}{2}\hat{Z}}$ by $u$ 
about $z$-axis,
and the C-NOT gate are realized 
in the one-way scheme as (a), (b), and (c) in Fig.~\ref{gates},
respectively.

\begin{figure}[htbp]
\begin{center}
\includegraphics[width=0.35\textwidth]{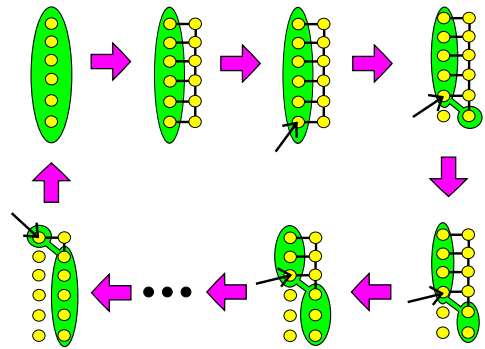}
\end{center}
\caption{(Color online.) The one-buffered implementation~\cite{Dawson} of
the one-way quantum computation.
The green ellipse represents the register state.
The black solid arrow represents the measurement.} 
\label{buffered}
\end{figure}

Let us also remind that there are two possibilities for
the implementation of
the one-way quantum computation.
One is that appeared in the original proposal~\cite{cluster} 
of the one-way quantum computation,
where the whole cluster state is created before the onset of 
adaptive measurements.
The other, which is called the ``one-buffered implementation"~\cite{Dawson},
is the repetition of the addition of a single column of the
cluster state to the register column and the measurement
of register qubits (see Fig.~\ref{buffered}).
We will adopt the one-buffered implementation.

\begin{figure}[htbp]
\begin{center}
\includegraphics[width=0.4\textwidth]{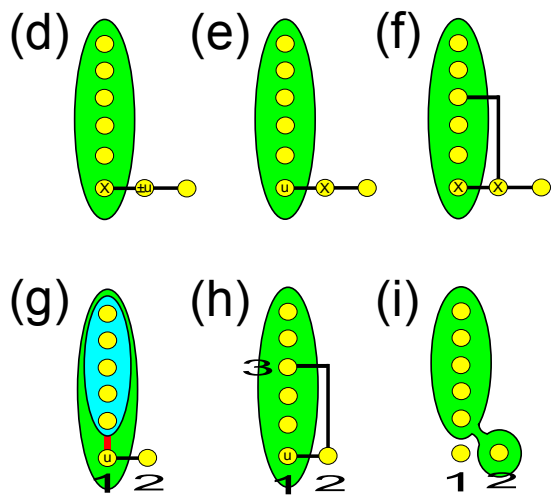}
\end{center}
\caption{(Color online.) 
(d): The register state $|\psi\rangle$ is represented by the green ellipse.
The rotation of the bottommost
qubit of $|\psi\rangle$ by $u$ about $x$-axis.
(e): The rotation of the bottommost
qubit of $|\psi\rangle$ by $u$ about $z$-axis.
(f): The C-NOT gate between the bottommost qubit and a qubit of $|\psi\rangle$.
(g): The red bond represents entanglement between the first qubit
and other register qubits (which are in the blue ellipse).
Processes (d) and (e) can be written as a combination of (g).
(h): The process (f) can be written as a combination of (g) and (h).
(i): The state after the measurement on the first qubit in (g).
} 
\label{mame}
\end{figure}

Let $|\psi\rangle$ be an $N$-qubit state, which is
considered as the quantum register.
We assume that one of the three operations, (a), (b), or (c),
in Fig.~\ref{gates}
is applied to $|\psi\rangle$ in the one-buffered implementation
as is shown in (d), (e), and (f)
in Fig.~\ref{mame}.
We are interested in the fidelity of these operations
assuming that a measurement is inaccurate.

\begin{figure}[htbp]
\begin{center}
\includegraphics[width=0.4\textwidth]{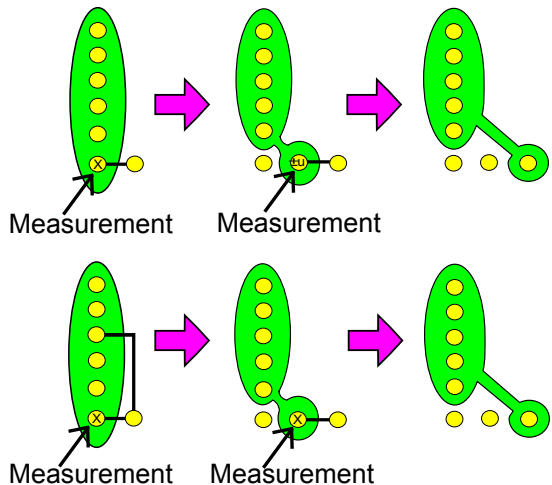}
\end{center}
\caption{(Color online.) 
Top line: The rotation of the bottommost qubit in $|\psi\rangle$
about $x$-axis.
The rotation about $z$-axis is given in a similar way.
Bottom line: The C-NOT gate between the bottommost qubit and
another qubit in $|\psi\rangle$.
} 
\label{mame2}
\end{figure}

It is easy to see that we have only to consider 
the fidelity of the process (g) in Fig.~\ref{mame}
for the study of (d), (e), and (f).
First, 
any of three operations, (d), (e), and (f) in Fig.~\ref{mame},
is a combination of
the two elementary processes (g) and (h) in Fig.~\ref{mame} 
(see Fig.~\ref{mame2}).
Therefore, the study of the fidelity of (d), (e), and
(f) are reduced to that of (g) and (h).
Second, in the process (h), 
the measurement on the first qubit [which is labeled as ``1"]
commutes with the C-Z interaction between the second qubit
and the third qubit [which are labeled as ``2" and ``3", respectively]. 
Therefore, the study of 
(h) is reduced to that of (g). 
In summary, we have only to consider the fidelity of 
the process (g) for our purpose.

\section{Main result}
Let us therefore calculate the fidelity of the process (g) in Fig.~\ref{mame}.
We assume that the measurement  
on the first qubit [which is labeled as ``1" in (g)]
is inaccurate in the sense that the direction to which the
qubit is projected is slightly deviated from the ideal one.
In other words,
the measurement
is not the ideal one $\{|u_+\rangle,|u_-\rangle\}$, where
\begin{eqnarray}
|u_\pm\rangle\equiv\frac{1}{\sqrt{2}}(|0\rangle\pm e^{-iu}|1\rangle),
\label{u}
\end{eqnarray}
but the slightly deviated one
$\{|\tilde u_+\rangle,|\tilde u_-\rangle\}$, where
\begin{eqnarray*}
|\tilde u_+\rangle&\equiv&
\cos\frac{\epsilon}{2}|u_+\rangle+e^{-i\delta}\sin\frac{\epsilon}{2}|u_-\rangle\nonumber\\
|\tilde u_-\rangle&\equiv&
\sin\frac{\epsilon}{2}|u_+\rangle-e^{-i\delta}\cos\frac{\epsilon}{2}|u_-\rangle.
\end{eqnarray*}
[If the measurement is done in the $\hat{X}$-basis,
we have only to put $u=0$ in Eq.~(\ref{u}).]
It is easy to see that the degree of 
the deviation is parametrized by $\epsilon$ and $\delta$:
$|\tilde u_+\rangle$ ($|\tilde u_-\rangle$)
is the vector obtained by rotating $|u_+\rangle$ ($|u_-\rangle$)
by $\epsilon$ about $z$-axis and by $\frac{\pi}{2}-\delta$ 
about $|u_+\rangle$-axis.
This kind of inaccuracy is ubiquitous in quantum physics.
For example, in the measurement model of von Neumann~\cite{Neumann},
the direction to which the primary state is projected
is deviated in this way if there is an inaccuracy in
the control of the coupling constant or the coupling time between
the primary system and the apparatus,
or if the projection measurement on the apparatus is inaccurate.

After the measurement of the first qubit in (g), 
the entanglement between the first qubit and other qubits is broken.
Then, (g) changes into (i) in Fig.~\ref{mame}. 
Let the register state after this measurement, i.e., the state
of qubits in the green ellipse
in (i), be $|\phi_{\epsilon,\delta}\rangle$.
If the measurement was accurate,
this is $|\phi_{0,0}\rangle$.
Then, we can show that
\begin{eqnarray}
F\equiv
{\mathbb E}\Big[\big|\langle\phi_{0,0}|\phi_{\epsilon,\delta}\rangle\big|^2\Big]
\le 1-S\sin^2\frac{\epsilon}{2}
\label{main}
\end{eqnarray}
where $\mathbb E[\cdot]$ means the average over all measurement histories,
\begin{eqnarray*}
S\equiv2[1-\mbox{Tr}(\hat{\rho}_1^2)]
\end{eqnarray*}
is the entanglement between the first qubit and other register qubits
[which is indicated by the red bond in (g)], 
and $\hat{\rho}_1\equiv\mbox{Tr}_1(|\psi\rangle\langle\psi|)$
is the reduced density operator of the first qubit ($\mbox{Tr}_1$
is the trace over all qubits 
except for the first qubit).
If the first qubit and other register qubits are not entangled,
$S=0$, whereas if they are maximally entangled, $S=1$.
Equation (\ref{main}) is our main result. 

{\it Proof of Eq.~(\ref{main})}:
Let us see (g) in Fig.~\ref{mame}.
The register state $|\psi\rangle$ (which is represented by the green ellipse) 
is written as
\begin{eqnarray*}
|\psi\rangle=\alpha|0\rangle_1\otimes|\eta_0\rangle_b
+\beta|1\rangle_1\otimes|\eta_1\rangle_b,
\end{eqnarray*}
where 
$|0\rangle_1$
and $|1\rangle_1$
are states of the first qubit [labeled as ``1" in (g)],
and $|\eta_0\rangle_b$
and $|\eta_1\rangle_b$
are states of other register qubits 
[represented by the blue ellipse in (g)].
$|\eta_0\rangle_b$
and $|\eta_1\rangle_b$
are not necessarily orthogonal with each other.
Let us add the second qubit $|+\rangle_2$ [labeled as ``2" in (g)]
to $|\psi\rangle$ and perform the C-Z
interaction between the first qubit and the second qubit:
\begin{eqnarray*}
|\psi\rangle\otimes|+\rangle_2
\to
\alpha|0\rangle_1\otimes|\eta_0\rangle_b\otimes|+\rangle_2
+\beta|1\rangle_1\otimes|\eta_1\rangle_b\otimes|-\rangle_2.
\end{eqnarray*}
As we have assumed,
the first qubit is measured in $\{|\tilde u_+\rangle,|\tilde u_-\rangle\}$.
Then, (g) changes into (i).
Let the states of the green ellipse in (i)
be $|\phi_{\epsilon,\delta}^\pm\rangle$ 
if the result of the measurement
is $\pm$, respectively.
By a straightforward calculation,
the probabilities $P_\pm$ of obtaining   
$|\phi_{\epsilon,\delta}^\pm\rangle$
are
\begin{eqnarray*}
P_\pm=\frac{1}{2}\big(1\pm \xi\sin\epsilon\cos\delta\big), 
\end{eqnarray*}
respectively,
where $\xi\equiv\mbox{Tr}(\hat{\rho}_1\hat{Z}_1)$. 
The fidelity for each output is
also calculated as
\begin{eqnarray*}
F_\pm\equiv
|\langle\phi_{0,0}^\pm|\phi_{\epsilon,\delta}^\pm\rangle|^2
=\frac{
1\pm \xi\sin\epsilon\cos\delta-(1-\xi^2)\sin^2\frac{\epsilon}{2}}
{2P_\pm},
\end{eqnarray*}
respectively,
and 
the mean fidelity 
is therefore 
\begin{eqnarray*}
F_+P_+ +F_-P_-
=1-(1-\xi^2)\sin^2\frac{\epsilon}{2}.
\end{eqnarray*}
Our goal, Eq.~(\ref{main}), is obtained by applying the relation
\begin{eqnarray*}
1-\mbox{Tr}^2(\hat{\rho}_1\hat{Z}_1)\ge S,
\end{eqnarray*}
which is shown as follows.
Let 
$\hat{\rho}_1=\lambda_0|\tau_0\rangle_1\langle\tau_0|
+\lambda_1|\tau_1\rangle_1\langle\tau_1|$,
where $\lambda_0\ge0$, $\lambda_1\ge 0$, $\lambda_0+\lambda_1=1$, and
\begin{eqnarray*}
|\tau_0\rangle_1&=&
\cos\frac{\mu}{2}|0\rangle_1+e^{-i\nu}\sin\frac{\mu}{2}|1\rangle_1\\
|\tau_1\rangle_1&=&
\sin\frac{\mu}{2}|0\rangle_1-e^{-i\nu}\cos\frac{\mu}{2}|1\rangle_1.
\end{eqnarray*}
Then, we obtain
$1-\mbox{Tr}^2(\hat{\rho}_1\hat{Z}_1)
=1-
(\lambda_0-\lambda_1)^2\cos^2\mu
\ge
1-(\lambda_0-\lambda_1)^2
=S$.

\section{Discussion}
\label{discussion}
If $S$ was always 0 during any quantum computation, Eq.~(\ref{main}) would be of no use.
However, in fact, $S$ often becomes very large
during a quantum computation.
For example, in Ref.~\cite{Jozsa}, it was shown that
if an $N$-qubit register state $|\psi\rangle$
is decomposed as the tensor product
of inseparable states $|\psi\rangle=\bigotimes_i|\psi_i\rangle$, 
at least one of these inseparable states $\{|\psi_i\rangle\}_i$
must have unboundedly increasing size during a quantum computation 
if the quantum computation offers an exponential speed-up over 
a classical one.
This result is not changed even if a weak entanglement is established
among $|\psi_i\rangle$'s.
Therefore,  
there is a high probability that
the measured qubit 
has sufficiently strong 
entanglement with other register qubits during a quantum computation.
Moreover, in Ref.~\cite{Ukena,visual}, it was shown that 
the register state has a superposition of
macroscopically distinct states
during the execution of Shor's factoring algorithm and Grover's search
algorithm. According to the result of Ref.~\cite{macroent}, 
a randomly chosen single qubit is strongly entangling with other qubits
with a high probability 
if the state has such a macroscopic superposition.
In short, $S$ is often very large in a quantum computation,
and therefore Eq.~(\ref{main}) offers a meaningful upper-bound
for the gate fidelity of the inaccurate one-way quantum computation.

The error model studied here is not an atypical one.
This type of error is indeed often considered in many studies of 
fault-tolerant quantum computations including Ref.~\cite{Dawson},
where the possibility of the fault-tolerant one-way quantum computation 
is shown.
Therefore, the effect of
our error is recoverable to some extent and the whole quantum
computation can be performed successfully.
However, as mentioned in Sec.~\ref{introduction},
the study of the stability of a bare one-way quantum computation
is very important.
This is where
our result can contribute.

In addition to the inaccurate measurement considered here,
there are many other possibilities of errors in the one-way quantum computation.
For example,
if the one-way quantum computation
is implemented with the discrete-variable linear-optics schemes~\cite{Nielsen2},
we must also consider the imperfection of the C-Z gate,
since, in this case, the entangling operation is not deterministic.
To consider other error models would lead to interesting generalizations
of the present work. It is left for a future study.

\acknowledgements
The author thanks T. Rudolph, M. S. Tame, A. Douglas K. Plato,
Y. Omar, and J. Kahn for discussion,
and the French Agence Nationale de la
Recherche (ANR) 
for support
under the grant StatQuant (JC07 07205763).





\begin{thebibliography}{00}

\bibitem{cluster}
R. Raussendorf and H. J. Briegel,
Phys. Rev. Lett. {\bf 86}, 5188 (2001).

\bibitem{Nielsen}
M. A. Nielsen and I. L. Chuang, 
{\it Quantum Computation and Quantum Information}
(Cambridge University Press, Cambridge, UK, 2000).

\bibitem{Furusawa}
M. Yukawa, R. Ukai, P. van Loock, and A. Furusawa,
Phys. Rev. A {\bf 78}, 012301 (2008);
X. Su, A. Tan, X. Jia, J. Zhang, C. Xie, and K. Peng,
Phys. Rev. Lett. {\bf 98}, 070502 (2007);
Y. Tokunaga, S. Kuwashiro, T. Yamamoto,
M. Koashi, and N. Imoto, Phys. Rev. Lett {\bf 100}, 210501 (2008);
G. Vallone, E. Pomarico, P. Mataloni, F. De Martini,
and V. Berardi, Phys. Rev. Lett {\bf 98}, 180502 (2007);
R. Ceccarelli, G. Vallone, F. De Martini, P. Mataloni,
and A. Cabello, Phys. Rev. Lett {\bf 103}, 160401 (2009).

\bibitem{Tame}
M. S. Tame, R. Prevedel, M. Paternostro, P. B\"ohi, M. S. Kim, 
and A. Zeilinger, 
Phys. Rev. Lett. {\bf 98}, 140501 (2007).

\bibitem{Walther}
P. Walther, K. J. Resch, T. Rudolph, E. Schenck, H. Weinfurther,
V. Vedral, M. Aspelmeyer, and A. Zeilinger,
Nature (London) {\bf 434}, 169 (2005).



\bibitem{Nest}
M. Van den Nest, A. Miyake, W. D\"ur, and H. Briegel,
Phys. Rev. Lett. {\bf 97}, 150504 (2006).
\bibitem{Gross}
D. Gross, S. T. Flammia, and J. Eisert,
Phys. Rev. Lett. {\bf 102}, 190501 (2009).
\bibitem{Bremner}
M. J. Bremner, C. Mora, and A. Winter, 
Phys. Rev. Lett. {\bf102}, 190502 (2009).

\bibitem{novelcluster}
D. Gross, J. Eisert, N. Schuch, and D. Perez-Garcia, Phys. Rev. A 
{\bf76}, 052315 (2007).

\bibitem{p}
A. Shimizu and T. Miyadera, Phys. Rev. Lett. {\bf89}, 270403 (2002).

\bibitem{macroent}
T. Morimae, Phys. Rev. A {\bf 81}, 010101(R) (2010).

\bibitem{tim}
T. Morimae, Phys. Rev. A {\bf81}, 022304 (2010).

\bibitem{magnon}
T. Morimae, A. Sugita, and A. Shimizu, Phys. Rev. A {\bf71}, 032317 (2005).

\bibitem{Dawson}
M. A. Nielsen and C. M. Dawson, Phys. Rev. A {\bf 71}, 042323 (2005).


\bibitem{Tame3}
M. S. Tame, M. Paternostro, M. S. Kim, and V. Vedral,
Phys. Rev. A {\bf 72}, 012319 (2005);
K. Kieling, T. Rudolph, and J. Eisert,
Phys. Rev. Lett. {\bf 99}, 130501 (2007);
K. Kieling, D. Gross, and J. Eisert, New. J. Phys. {\bf 9}, 200 (2007);
P. P. Rohde, S. D. Barrett, New J. Phys. {\bf 9}, 198 (2007);
D. Jennings, A. Dragan, S. D. Barrett,
S. D. Bartlett, and T. Rudolph,
Phys. Rev. A {\bf 80}, 032328 (2009).

\bibitem{other}
To consider other resource states,
such as those studied in Ref.~\cite{novelcluster}, would be an
interesting subject of a future study.


\bibitem{Jozsa}
R. Jozsa and N. Linden, Proc. R. Soc. Lond. A {\bf 459}, 2011 (2003).

\bibitem{Ukena}
A. Ukena and A. Shimizu, Phys. Rev. A {\bf 69}, 022301 (2004).

\bibitem{visual}
T. Morimae and A. Shimizu, Phys. Rev. A {\bf74}, 052111 (2006).

\bibitem{debbie}
P. Aliferis and D. W. Leung, Phys. Rev. A {\bf 70}, 062314 (2004).



\bibitem{Neumann}
J. von Neumann, {\it Mathematical Foundations
of Quantum Mechanics} (Princeton Univ. Press, Princeton, 1955). 





\bibitem{Nielsen2}
M. A. Nielsen, Phys. Rev. Lett. {\bf 93}, 040503 (2004).

\end{thebibliography}
\end{document}